\documentclass[twocolumn,aps,prl,amsmath,amssymb,showpacs,preprintnumbers]{revtex4}
\begin{document}

\preprint{AEI--2005--115}
\preprint{gr-qc/0506128}

\title{Nonsingular Black Holes and Degrees of Freedom in Quantum Gravity}

\author{Martin Bojowald}
\email{mabo@aei.mpg.de}
\affiliation{Max-Planck-Institut f\"ur Gravitationsphysik, Albert-Einstein-Institut,\\
Am M\"uhlenberg 1, D-14476 Potsdam, Germany}


\newcommand{\lP}{\ell_{\mathrm P}}
\newcommand{\vp}{\varphi}
\newcommand{\vt}{\vartheta}

\newcommand{\md}{{\mathrm{d}}}
\newcommand{\tr}{\mathop{\mathrm{tr}}}
\newcommand{\sgn}{\mathop{\mathrm{sgn}}}

\newcommand*{\R}{{\mathbb R}}
\newcommand*{\N}{{\mathbb N}}
\newcommand*{\Z}{{\mathbb Z}}

\begin{abstract}
  Spherically symmetric space-times provide many examples for
  interesting black hole solutions, which classically are all
  singular.  Following a general program, space-like singularities in
  spherically symmetric quantum geometry, as well as other
  inhomogeneous models, are shown to be absent.  Moreover, one sees
  how the classical reduction from infinitely many kinematical degrees
  of freedom to only one physical one, the mass, can arise, where
  aspects of quantum cosmology such as the problem of initial
  conditions play a role.
\end{abstract}

\maketitle

One of the main issues to be addressed by quantum gravity is the
singularity problem of general relativity. While the classical theory
is very successful in describing space-time on scales which can be
probed today, it is incomplete because it predicts the generic presence
of singularities: boundaries of space-time which can be reached by
observers in a finite amount of time, but at which point the theory
becomes inapplicable. Usually, curvature or energy densities and tidal
forces diverge there, implying unphysical conditions.

One explanation is that the picture of a smooth space-time underlying
the classical theory is only appropriate at large scales, while at
small scales the structure is discrete. This has indeed been
substantiated by looking at cosmological models in loop quantum
gravity \cite{Rev}, where the discrete quantum geometry has
been shown to remove singularities \cite{Sing}. Even at the classical
level there are indications for a break-down of the smooth picture at
small scales: The BKL scenario \cite{BKL} provides a scheme for the
general approach to a classical singularity by considering dominant
contributions to the field equations in this limit. It turns out that
only terms with time-derivatives remain such that spatial
points decouple and their geometries can be described by the most
general homogeneous model. This Bianchi IX model is
classically chaotic \cite{NumSing} and so geometries in different points are
completely unrelated, implying a complicated classical singularity
with structure at arbitrarily small scales.

Since the removal of singularities in loop quantum cosmology applies
for all homogeneous models \cite{HomCosmo,Spin}, in particular the
Bianchi IX model, one can combine this result with the BKL picture and
expect that all singularities are removed by quantum geometry.
However, the BKL picture has not been proven classically, and the
above argument would require it to hold even in quantum gravity. In
particular the latter point is questionable because, for one, already
the approach to a single Bianchi IX singularity \cite{NumSing} in
quantum cosmology is modified, removing the classical chaos
\cite{NonChaos,ChaosLQC}. It is thus necessary to study inhomogeneous
models in loop quantum gravity and look at the singularity issue
without assuming homogeneity. If this is possible, one can also test
the validity of the BKL picture in the quantum context. This is what
we will do here in the case of spherical symmetry, which is not only
the simplest inhomogeneous situation but also allows interpretations
for black holes.  The same methods apply to other models which do have
local degrees of freedom, providing the first demonstration of the
absence of singularities in inhomogeneous quantum gravity.

\paragraph{Singularities.} The main problem caused by a
singularity is the fact that it presents a boundary to physical
evolution. In order to see whether it persists in quantum gravity,
then, the following steps have to be performed. This has to be done in
a manner which is independent of coordinate or other gauge choices,
and only potential simplifications resulting from the symmetry
reduction should be used. One first has to locate classical
singularities on the phase space of physical fields, the spatial
metric $q_{ab}$ and extrinsic curvature related to $\dot{q}_{ab}$.
Conditions to specify the {\em singular part of phase space} must be
chosen such that any solution to the theory, which is a trajectory on
phase space, intersects this singular part exactly when it develops a
singularity. The solution space is in general quite
complicated to study, but one can select a variable $T$ on phase space
which is transversal to the singular part, a {\em local internal time}
rather than coordinate time, i.e.\ which fulfills $T=0$ in a
neighborhood around zero exactly at the singular part. Finally, one
needs to write down the quantum evolution of geometry in the local
internal time and check whether or not it stops at $T=0$.
If one can find a $T$ such that the quantum evolution does not stop
anywhere, the quantum system is non-singular. This is the analog of
the classical notion of space-time completeness.

We illustrate this scheme with isotropic cosmology where the phase
space is 2-dimensional with the scale factor $a$ (the spatial radius
of the universe) and its time derivative. Singularities occur only if
the scale factor vanishes such that $a=0$ specifies the singular part.
An obvious local internal time (which in this case is global) is given
by $T=a$, or with a slight modification the triad variable $p$ with
$|p|=a^2$ and $\sgn p$ being the orientation of space. Using this
variable makes no difference classically, but is important in quantum
geometry where triads are basic variables. At the quantum level one
can then first note that operators for $p^{-1}$ are finite
\cite{InvScale}, indicating already that curvatures and energy
densities do not diverge, and most importantly that the quantum
evolution is given by a difference equation for the wave function in
$p$ which {\em does not stop at $p=0$} \cite{Sing}. Thus, there is no
singularity in isotropic loop quantum cosmology.

\paragraph{Spherical symmetry.} The case of interest here is spherical
symmetry, where the kinematical phase space on which we have to locate
singularities is infinite-dimensional and spanned by the metric
components in
\begin{equation}
 \md s^2 = q_x(x) \md x^2+q_{\Omega}(x)\md\Omega^2
\end{equation}
(in polar coordinates) and their time derivatives. As an example we
can look at the Schwarzschild solution for a black hole of mass $M$,
$q_x=(1-2M/x)^{-1}$, $q_{\Omega}=x^2$.  The singularity is reached for
$x=0$, at which point both metric coefficients are zero. The question
then arises which one, or both, of them must be zero as a condition
for a singularity. It turns out that $q_{\Omega}$ is zero only at the
singularity, while $q_x$ can also become zero elsewhere, i.e.\ at the
horizon $x=2M$, when one chooses a different gauge (e.g.\ with
homogeneous coordinates in the interior).  This point illustrates why
gauge independence is essential in answering the singularity problem:
even the very first step, finding where singularities would develop,
depends on it. In fact, in this case we can choose our coordinates $x$
and $t$ at will, which affects the form of $q_x$ and points where it
can be zero. In spherical symmetry, however, the fact that
$q_{\Omega}=0$ at the singularity is unaffected (even though, of
course, $q_{\Omega}$ can change as a function of $x$ when we change
coordinates).

We can now consider a spatial slice which locally, around a point
$x_0$, approaches the classical singularity such that
$q_{\Omega}(x_0)\to0$. The above discussion shows that
$T=q_{\Omega}(x_0)$ is a good local internal time, which completes
setting up the problem from the classical side. It now remains to
formulate the quantum evolution in local internal time and check if it
stops at $T=0$.

\paragraph{Quantum geometry.} Again we first transform to triad
variables $|E^x|=q_{\Omega}$ and $E^{\vp}=\sqrt{q_xq_{\Omega}}$ which
become basic operators in quantum geometry. The (local)
orientation of space around a point $x_0$ is now given by $\sgn
E^x(x_0)$ where $E^x$ unlike $E^{\vp}$ can take both signs. Moreover,
the discussion in metric variables shows that $T=E^x(x_0)$ is our
local internal time such that the situation, so far, is analogous to
that in the isotropic case: triad variables lead to a local internal
time which takes values at both sides of the classical singularity,
$T=0$ defining a manifold in superspace rather than at the boundary.
It is important to note that the introduction of triad variables was
seen as a necessary step toward a background independent quantization.
Now it turns out that this also changes the singularity structure on
phase space in a way which was important for removing cosmological
singularities. Nevertheless, even though the singularity is now
located in superspace, the classical evolution still stops there and
is not able to connect from positive to negative $T$. This still has
to be checked by the quantum evolution, the most crucial point.

Quantum evolution
follows from the Hamiltonian constraint operator acting on
states in the form of a lattice model with basis \cite{SphSymm}
$|\vec{k},\vec{\mu}\rangle:=${\unitlength=0.2mm\begin{picture}(200,10)(0,2)
    \put(10,5){\line(1,0){180}} \put(50,5){\circle*{5}}
    \put(100,5){\circle*{5}} \put(150,5){\circle*{5}}
 \put(50,-3){\makebox(0,0){$\cdots$}}
 \put(100,-3){\makebox(0,0){$\mu_n$}}
 \put(150,-3){\makebox(0,0){$\cdots$}}
 \put(25,10){\makebox(0,0){$\cdots$}}
 \put(75,12){\makebox(0,0){$\scriptstyle k_n$}}
 \put(125,12){\makebox(0,0){$\scriptstyle k_{n+1}$}}
 \put(175,10){\makebox(0,0){$\cdots$}}
\end{picture}}
where the integer labels $k_e$ on edges are eigenvalues of the
operator $\hat{E}^x$ and the positive real labels $\mu(v)$ at vertices
those of $\hat{E}^{\vp}$. Positions of vertices do not refer to a
background space, and the lattice model represents the continuum
theory. The constraint then acts by \cite{SphSymmHam}
\begin{center}
$\hat{H}[N]$ 
{\unitlength=0.2mm
\begin{picture}(100,10)(50,2)
 \put(50,5){\line(1,0){100}}
 \put(100,5){\circle*{5}}
 \put(75,12){\makebox(0,0){$\scriptstyle k_-$}}
 \put(125,12){\makebox(0,0){$\scriptstyle k_+$}}
\end{picture}}
 $=\sum_v N(v)\bigl(\hat{C}_0(k)$~{\unitlength=0.2mm
\begin{picture}(100,10)(50,2)
 \put(50,5){\line(1,0){100}}
 \put(100,5){\circle*{5}}
 \put(75,12){\makebox(0,0){$\scriptstyle k_-$}}
 \put(125,12){\makebox(0,0){$\scriptstyle k_+$}}
\end{picture}} $+\hat{C}_{R+}(k)$~{\unitlength=0.2mm
\begin{picture}(100,10)(50,2)
 \put(50,5){\line(1,0){100}}
 \put(100,5){\circle*{5}}
 \put(75,12){\makebox(0,0){$\scriptstyle k_-$}}
 \put(125,12){\makebox(0,0){$\scriptstyle k_++2$}}
\end{picture}} $+\hat{C}_{R-}(k)$~{\unitlength=0.2mm
\begin{picture}(100,10)(50,2)
 \put(50,5){\line(1,0){100}}
 \put(100,5){\circle*{5}}
 \put(75,12){\makebox(0,0){$\scriptstyle k_-$}}
 \put(125,12){\makebox(0,0){$\scriptstyle k_+-2$}}
\end{picture}} $+\hat{C}_{L+}(k)$~{\unitlength=0.2mm
\begin{picture}(100,10)(50,2)
 \put(50,5){\line(1,0){100}}
 \put(100,5){\circle*{5}}
 \put(75,12){\makebox(0,0){$\scriptstyle k_-+2$}}
 \put(125,12){\makebox(0,0){$\scriptstyle k_+$}}
\end{picture}} $+\hat{C}_{L-}(k)$~{\unitlength=0.2mm
\begin{picture}(100,10)(50,2)
 \put(50,5){\line(1,0){100}}
 \put(100,5){\circle*{5}}
 \put(75,12){\makebox(0,0){$\scriptstyle k_--2$}}
 \put(125,12){\makebox(0,0){$\scriptstyle k_+$}}
\end{picture}} $+\cdots\bigr)$
\end{center}
summing over all vertices of the lattice, the dots indicating
further terms such as a matter Hamiltonian whose detailed form is not
important here. The known coefficients
$\hat{C}_I(k)=C_I(k)\hat{C}_I$ consist of functions $C_I(k)$ of the edge
labels and operators $\hat{C}_I$ acting only on the dependence on vertex
labels $\mu$.  A general state is now a superposition $|\psi\rangle=
\sum_{\vec{k},\vec{\mu}}\psi(\vec{k}, \vec{\mu}) |\vec{k},\vec{\mu}\rangle$ 
whose coefficients $\psi(\vec{k},\vec{\mu})$ define the state in the
triad representation. The constraint $\hat{H}[N]|\psi\rangle=0$
has to hold true for all functions $N$ with independent values $N(v)$,
giving one equation for each vertex which in the triad representation
takes the form
\begin{eqnarray*}
 && \hat{C}_0(k)\psi(k_-,k_+)+
 \hat{C}_{R+}(k)\psi(k_-,k_+-2)\\
 &&+\hat{C}_{R-}(k)\psi(k_-,k_++2)
 +\hat{C}_{L+}(k)\psi(k_--2,k_+)\\
 &&+\hat{C}_{L-}(k)\psi(k_-+2,k_+)+\cdots =0
\end{eqnarray*}
of a difference equation, where we have suppressed the vertex labels
on which the $\hat{C}_I$ act and unchanged $k$.

We now solve this set of equations with initial and boundary values
for the wave function. To define a solution scheme we proceed
iteratively from vertex to vertex, starting at one side $\partial$ of
the lattice. We assume that the boundary values for all
$\mu_{\partial}$ and $k_+(\partial)=:k_-$ of the wave function as well
as values for large positive $k_e=k_0$ and $k_0-1$ at all edges $e$
are given, which means that we have specified the initial situation,
e.g.\ by a semiclassical state specifying the initial slice far away
from the singularity.  The equation can then be solved for
$\hat{C}_{R+}\psi(k_-,k_+-2)$ in terms of values of the wave function
specified by the initial conditions. This brings us one step further
because we now have information about the wave function at $k_+-2$ for
a smaller edge label (our local internal time) evolving toward the
classical singularity.

Next, we have to know how to find $\psi$ from its image under
$\hat{C}_{R+}$. This can be done by specifying conditions for the wave
function at small $\mu$ (which is not in the singular part of
minisuperspace but represents an ordinary boundary) and happens in
exactly the same way as in homogeneous models \cite{HomCosmo}.  Before
continuing, we notice that this indicates the presence of aspects of
the BKL picture in quantum gravity. However, we still have to try to
evolve through the classical singularity, i.e.\ $k_e=0$, which will be
the main test.  One crucial difference to cosmological models is that
the coefficients $\hat{C}_I(k)$ are not only functions of the local
internal time, $k_+$, studied in the iteration but also of neighboring
labels such as $k_-$ which do not take part in this difference
equation but the dependence on which has been determined in iteration
steps for previous vertices.  This is clearly a new feature coming
from the inhomogeneous context, and it has a bearing on the
singularity issue.

Singularities are removed if the difference equation determines the
wave function everywhere on minisuperspace once initial and boundary
conditions have been chosen away from classical singularities. The
simplest realization is by a difference equation with non-zero
coefficients everywhere. However, this is not automatically the case
with an equation coming from a general construction of the Hamiltonian
constraint, and so has to be checked explicitly. Here, it turns out
\cite{SphSymmHam} that a symmetric constraint indeed leads to non-zero
functions $C_I(k)$ which then will not pose a problem to the
evolution. All values of the wave function, at positive as well as
negative $k$, are determined uniquely by the difference equations and
chosen initial and boundary values. The evolution thus continues
through the classical singularity at zero $k$: {\em there is no
quantum singularity}.
Other quantization choices can lead to quantum singularities,
providing selection criteria to formulate the quantum theory with
implications also for the full framework.

\paragraph{Consequences.} We have shown that the same mechanism as in
homogeneous models contributes to the removal of spherically symmetric
classical singularities. Key features are that densitized triads as
basic variables in quantum geometry provide us with a local internal
time taking values at {\em two sides of the classical singularity},
combined with a quantum evolution that {\em connects both sides}.  No
new ingredients are necessary for inhomogeneous singularities, only an
application of the general scheme to the new and more complicated
situation.

As in cosmological models the argument applies {\em only to space-like
singularities} such as the Schwarzschild one. The reason is that we
evolve a spatial slice toward the classical singularity and test
whether it will stop. A time-like or null singularity would require a
different mechanism which is not known at present. Thus, cases like
negative mass solutions seem to remain singular, which is a welcome
property helping to rule out unwanted solutions leading to
instability \cite{SingValue}.

This scenario and its form of difference equations does not only apply
to vacuum black holes but also to spherically symmetric matter
systems. In such a case, there would be new labels for matter fields,
and a contribution to the constraint from the matter Hamiltonian. As
this does not change the structure of the difference equation, the
same conclusions apply. Moreover, models for Einstein--Rosen waves
have a similar structure just with a new vertex label. Also in this
case, with or without matter fields, the analysis goes through such
that the absence of singularities can be demonstrated even in
situations with local gravitational degrees of freedom.

There are differences between homogeneous models and these
inhomogeneous cases, and the inhomogeneous analysis is much more
non-trivial. In homogeneous models there are several ambiguities in
the constraint operator, and several choices lead to non-singular
evolution. In more complicated situations such as those studied here,
not all options remain available. In particular, we had to use a
symmetric ordering of the constraint in order to have non-vanishing
coefficients of the difference equation. In homogeneous models one can
also work with a version whose coefficients vanish right at the
singularity. The evolution then still continues since the value at
the classical singularity simply decouples and does not play a role
for the evolution. Instead, one can use the behavior to find dynamical
initial conditions \cite{DynIn}. This is also possible here for
evolution in local internal time, but then the decoupled value at
$k_-=0$ is not determined and in general needed for the wave function
at other values of $k_+$. The inhomogeneous evolution would thus break
down, and this choice of constraint is ruled out.

There is a difference in the constraint operator we used compared to a
common expression in the full theory \cite{QSDI}. This issue is
visible only in inhomogeneous models, and consists in whether or not
the constraint creates new edges and vertices, or just changes labels
of existing ones. We chose the second possibility, which has already
been considered as a modification in the full theory
\cite{S:ClassLim}. There, it can better explain the presence of
correlations at an intuitive level, but makes checking anomaly-freedom
more complicated. The main problem of an anomalous quantization would
be that too many states could be removed when imposing the
constraints, leaving not enough physical solutions. This issue can be
checked here with the constraint we used. If there is no matter field
present we expect just one physical degree of freedom, the
Schwarzschild mass $M$. In our solution scheme we started with a
boundary state $\psi_{\partial}$ corresponding to this degree of
freedom, and with this state being free it is already clear that we do
not lose too many states. It is even possible to check whether or not
the number of independent physical solutions is correct, i.e.\ not too
large either. In the iteration we solve one difference equation for
$\psi$ at each vertex, such that any freedom here would provide new
quantum degrees of freedom. Since the difference equation for $\psi$
has the same form as that in homogeneous loop quantum cosmology, the
{\em number of quantum degrees of freedom is formally related to the
initial value problem of quantum cosmology}.  A possible physical
meaning is to be checked in explicit examples.

In the isotropic case there are indeed dynamical initial conditions
following from the dynamical law \cite{DynIn,Essay} which, if realized
in our context, would imply that solutions for $\psi$ are unique and
the mass is the only quantum degree of freedom. However, these
conditions rely on the fact that leading coefficients of the
difference equation can vanish, which we have ruled out for
inhomogeneous models. Moreover, the uniqueness of a quantum
cosmological wave function depends on the pre-classicality condition
of \cite{DynIn}.  Other mechanisms to select unique cosmological
solutions are thus needed, such as from observables or the physical
inner product \cite{IsoSpinFoam}. This issue is quite complicated for
difference equations in particular in anisotropic models
\cite{GenFuncBI}.  Nonetheless, a simple counting of free variables
supports the connection to initial conditions: The vacuum spherically
symmetric case has difference equations in three independent
variables, an edge label $k$ and two neighboring vertex labels $\mu$.
Homogeneous loop quantum cosmology gives rise to an equation of
similar structure and also three variables, so if we assume that there
is a mechanism for a unique solution it will also apply to black holes
of a given mass.  Adding matter fields (or more gravitational freedom
as in Einstein--Rosen) increases the number of independent variables to
five in inhomogeneous models ({\em two} new vertex labels) as opposed
to four in homogeneous matter models. The type of difference equations
thus agrees in homogeneous and inhomogeneous models in vacuum, but not
when local degrees of freedom are present.

Thus, the structure of the Hamiltonian constraint equation from loop
quantum gravity can potentially provide explanations for issues as
diverse as the singularity problem in cosmology and black hole
physics, initial conditions in quantum cosmology, the semiclassical
limit and issue of quantum degrees of freedom. We emphasize that many
of these connections still have to be checked in generality.  Still,
such connections between seemingly unrelated issues in quantum gravity
can be seen as support for the internal consistency of the whole
theory and, hopefully, provide guidance for future developments.

We can finally come back to the approach to a classical singularity
and the BKL picture. Our results here do not rely on an extension of
the BKL picture to the quantum situation.  First of all, the situation
is conceptually different because evolution is now studied for a wave
function in local internal time $T$, rather than the spatial metric in
coordinate time.  Nevertheless, at first sight a similar picture
arises here from the quantum equation: as used in the previous
arguments, the equations can be reduced to ordinary difference
equations in $T$, where neighboring edges just contribute via an
inhomogeneity of the difference equation.  The inhomogeneous
situation, however, does play an important role right at the classical
singularity where some versions which would be allowed in homogeneous
models are ruled out.

Given that the techniques necessary for the quantum theory are similar
to lattice models, it is easy to implement them in numerical quantum
gravity. This opens the door to numerical investigations of many
problems that are still actively pursued in classical gravity
\cite{NumSing}, such as the approach to classical singularities and
the issue of gravitational collapse and naked singularities. This
requires studying horizons in addition to classical singularities,
which can also be done at the quantum level \cite{Horizon}.  As we
have seen, there are many non-trivial quantum effects which play
together in just the right way to ensure the absence of singularities,
which has prospects for other effects in the physics of black holes
\cite{BHPara}.

\end{document}